\documentclass[english,preprint]{aastex}
\usepackage[T1]{fontenc}
\makeatletter

\providecommand{\tabularnewline}{\\}

\usepackage{babel}
\makeatother
\begin{document}

\title{Probing the magnetic field with molecular ion spectra.}

\author{Martin Houde \altaffilmark{1,2},\email{houde@ulu.submm.caltech.edu}
Pierre Bastien\altaffilmark{2}, Ruisheng Peng\altaffilmark{1}, Thomas
G. Phillips \altaffilmark{3} \and Hiroshige Yoshida \altaffilmark{1}}

\altaffiltext{1}{Caltech Submillimeter Observatory, 111 Nowelo Street, Hilo, HI 96720}

\altaffiltext{2}{Département de Physique, Université de Montréal, Montréal, Québec H3C 3J7, Canada}

\altaffiltext{3}{California Institute of Technology, Pasadena, CA 91125}

\begin{abstract}
Observations of the effect of the magnetic field on its environment
are usually achieved with techniques which rely on the interaction
with the spin of the particles under study. Because of the relative
weakness of this effect, extraction of the field characteristics proves
to be a most challenging task. We take a totally different approach
to the problem and show that the manifestation of the magnetic field
can be directly observed by means of a comparison of the spectra of
molecular ions with those of neutral molecules. This takes advantage
of the strong cyclotron interaction between the ions and the field,
but requires the presence of flows or turbulent motion in the gas.
We compare our theory to data obtained on the OMC1, OMC2, OMC3 and
DR21OH molecular clouds. 
\end{abstract}


\section{Introduction.}

The suspected effects of the magnetic field in the process of star
formation are well documented in the literature. They may occur during
the initial collapse of clouds, by providing frictional support through
collisions between ions and neutrals, or through the establishment
of a pressure associated with long wavelength magnetohydrodynamic
waves arising from small scale fluctuations in the field, or again
as source of magnetohydrodynamic forces at the center of the molecular
outflow phenomenon \citep{Shu et al. 1987,Mouschovias 1991a,Mouschovias 1991b,Crutcher 1993}.
The need for accurate measurements of the magnetic field and its effects
on the environment cannot be understated. This is, however, a formidable
task. Up to now, the observed manifestations of the magnetic field,
in star forming regions, consist of Zeeman line broadening measurements
(which are best carried out at low frequencies, due to the increasing
dominance of the Doppler width over Zeeman splitting with increasing
frequency) and in the magnetic alignment of the dust grains known
to exist in molecular clouds.

Despite the weakness of the interaction between the field and the
observed molecular species and the numerous difficulties inherent
to the technique, Zeeman measurements provide us with the only way
of measuring the intensity of the magnetic field (more specifically
its component normal to the plane of the sky \citep{Crutcher 1993,Crutcher et al. 1999}).
On the other hand, a measure of the degree of alignment of the grains
can be obtained from the level of linear polarization detected from
continuum emission emanating from molecular clouds. The polarization
is believed to be caused by anisotropic emission from dust grains.
Recent studies \citep{Hildebrand 1999,Draine 1996} reveal that the
intensity of the magnetic field has very little to do with the level
of polarization induced by the dust. The field is needed to align
the grains and the level of polarization is basically a function of
the spin imparted to the individual grains by some agent, most likely
the radiation field. The thermally generated spin is known to be too
small to explain the observed effect \citep{Draine 1996}.

In this paper, we present theoretical and observational evidence for
a different effect which demonstrates the presence of the magnetic
field. Its signature resides in the line profiles of molecular ion
species at millimeter and submillimeter wavelengths. It will be shown
that, under the right conditions, even a weak field ($\sim10$~$\mu$G)
will cause ions to exhibit narrower profiles and a suppression of
high velocity wings when compared to the lines of neutral species.

There are of course several possible effects that could confuse the
issue, including optical depth differences between species and lack
of spatial coexistence. We deal with such problems in part in this
paper and in more depth in a subsequent one.

\section{Ion spectra versus neutral spectra.\label{sec:analysis}}

Probably the first thing we should point out is that the lines observed
in molecular clouds are usually many times broader than their thermal
width. It is generally assumed that this is due to the presence of
turbulence in the interstellar medium \citep{Zuckerman 1974,Falgarone 1990}.
This will be the point of view adopted here. Turbulence is characterized
by vortex streaming, the motions taking place in eddies of different
sizes \citep{Tennekes 1972}. In what follows, we will concentrate
on a small portion of a given eddy so that the region considered is
small enough that motions of the local flow can be approximated as
being linear, with the understanding that on a larger scale the motion
exhibits vorticity. This analysis can be applied without approximation
to cases involving linear flows (e.g., jets, outflows, \ldots{}).

The effect to be discussed is the tendency for ions to be forced into
gyromagnetic motion about the magnetic field direction, rather than
following the streaming flow of the general neutral motion. It is
necessary to discuss the ion-neutral collisions and we start with
a simplified approach. We will assume that we are dealing with a weakly
ionized plasma, where the neutral flow is mainly composed of molecular
hydrogen; we therefore assume a molecular mass number of $A_{n}=2.3$
for its constituents.

\subsection{Approximate solution - head on collisions.}

The equation of motion of a single ion subjected to a flow of neutral
particles in the presence of a magnetic field is given by:

\begin{equation}
\frac{d\mathbf{v}}{dt}=\frac{e}{m_{i}c}\,\mathbf{v}\times\mathbf{B}+\mathbf{F}_{c}\,\,,\label{eq:motion}\end{equation}

where $m_{i}$, $\mathbf{v}$, $\mathbf{B}$ are the mass of the ion,
its velocity and the magnetic field. $\mathbf{F}_{c}$ is the force
(per unit mass) on the ion due to the collisions with neutrals. The
nature of the interaction during a collision can be quite complicated
and how the momentum is transfered between the particles needs to
be handled carefully. But in order to get an idea of the behavior
of the ion we will initially assume that all the collisions are perfectly
elastic and {}``head on''. Accordingly, the force of interaction
can be approximated by:

\begin{equation}
\mathbf{F}_{c}=-2\frac{\mu}{m_{i}}\left(\mathbf{v}-\mathbf{v}^{n}\right)\sum_{m=-\infty}^{\infty}\delta\left(t-\tau_{m}\right)\,\,,\label{eq:force}\end{equation}

where $\mu$, $\mathbf{v}^{n}$, $\tau_{m}$ are the reduced mass,
the neutral flow velocity and the different times at which collisions
randomly occur. It will make things easier if we break up the velocities
into two components, one parallel to the mean magnetic field ($\mathbf{v}_{\Vert}$
and $\mathbf{v}_{\Vert}^{n}$) and the other perpendicular ($\mathbf{v}_{\bot}$
and $\mathbf{v}_{\bot}^{n}$). In steady state conditions, it can
be shown that the mean and variance of the velocity components are
given by (see the appendix for details):

\begin{eqnarray}
\left\langle \mathbf{v}_{\Vert}\right\rangle  & = & \left\langle \mathbf{v}_{\Vert}^{n}\right\rangle \label{eq:avmz}\\
\left\langle \mathbf{v}_{\bot}\right\rangle  & = & \frac{\left\langle \mathbf{v}_{\bot}^{n}\right\rangle +\left\langle \omega_{r}\right\rangle ^{-1}\left[\left\langle \mathbf{v}_{\bot}^{n}\right\rangle \times\left\langle \overrightarrow{\omega_{g}}\right\rangle \right]}{1+\left(\frac{\left\langle \overrightarrow{\omega_{g}}\right\rangle }{\left\langle \omega_{r}\right\rangle }\right)^{2}}\label{eq:avmp}\\
\sigma_{\Vert}^{2} & = & \frac{\left[\sigma_{\Vert}^{n}\right]^{2}}{\left[\frac{m_{i}}{\mu}-1\right]}\label{eq:asigz}\\
\sigma_{\bot}^{2} & = & \frac{\left\langle \left|\mathbf{v}_{\bot}^{n}\right|^{2}\right\rangle -\left\langle \mathbf{v}_{\bot}\right\rangle ^{2}}{\left[\frac{m_{i}}{\mu}-1\right]}\label{eq:asigp}\\
\sigma_{T}^{2} & = & \sigma_{\Vert}^{2}+\sigma_{\bot}^{2}\label{eq:asigt}\end{eqnarray}

with

\begin{eqnarray}
\left\langle \overrightarrow{\omega_{g}}\right\rangle  & = & \frac{e\left\langle \mathbf{B}\right\rangle }{m_{i}c}\label{eq:gyrofreq}\\
\left\langle \omega_{r}\right\rangle  & = & 2\frac{\mu}{m_{i}}\nu_{c}\label{eq:awrelax}\end{eqnarray}

$\left\langle \omega_{r}\right\rangle $ is what we will call, for
reasons to be discussed later, the relaxation rate. $\left\langle \overrightarrow{\omega_{g}}\right\rangle $,
$\nu_{c}$ are the mean ion gyrofrequency vector and the (mean) collision
rate. 

Equation (\ref{eq:avmp}) represents the drift that the ion can possess
in relation to the neutral flow and it is at the heart of the effect
that we are now studying. Equation (\ref{eq:asigp}) gives a measure
of the gyration amplitude of the ion around a guiding center in the
region occupied by the magnetic field. We have also allowed for an
inherent velocity dispersion in the neutral flow as can be seen by
the presence of the term $\sigma_{\Vert}^{n}$ and the fact that $\left\langle \left|\mathbf{v}_{\bot}^{n}\right|^{2}\right\rangle \neq\left\langle \mathbf{v}_{\bot}^{n}\right\rangle ^{2}$.

From equations (\ref{eq:avmz})-(\ref{eq:asigt}) it can be deduced
that, as could be expected, the ion completely follows the flow when
the latter is aligned with the field ($\mathbf{v}_{\bot}^{n}=0$).
More interesting, however, is the ion behavior when the neutral flow
is perpendicular to the field ($\mathbf{v}_{\Vert}^{n}=0$). In such
cases, the following observations can be made:

\begin{itemize}
\item for weak field intensities ($\left\langle \omega_{g}\right\rangle \ll\left\langle \omega_{r}\right\rangle $)
the ion follows the flow as $\left\langle \mathbf{v}_{\bot}\right\rangle \sim\left\langle \mathbf{v}_{\bot}^{n}\right\rangle $
\item as the field gains in strength the ion starts to drift in a direction
perpendicular to both the flow and the field until we get to the point
where $\left\langle \mathbf{v}_{\bot}\right\rangle \sim0$ when the
field reaches high intensities ($\left\langle \omega_{g}\right\rangle \gg\left\langle \omega_{r}\right\rangle $).
The ion is then basically trapped in the field and its mean square
velocity can be evaluated solely with equation (\ref{eq:asigp}),
$\sigma_{\bot}$ is found to be smaller than the flow velocity by
a factor of a few. For example, if we choose $A_{i}=29$ and $A_{n}=2.3$
for the ion and neutral molecular mass numbers we find this factor
to be $\sim3.5$. That is, \emph{an ion would have on average an effective
velocity at least $\sim3.5$ times smaller than that of a neutral
molecule of the same mass}.
\end{itemize}
This last observation has important implications. Let us assume that
an observed emission line from a neutral molecular species is composed
of contributions from a family of flows (or eddies) of different velocities.
We can then infer that \emph{for regions inhabited with a strong enough
magnetic field which is, on average, not aligned with the local flow(s),
we expect ionic lines to exhibit narrower profiles and a suppression
of high velocity wings when compared to neutral lines}.

Obviously, we cannot expect all the flows to be perpendicular to the
mean magnetic field. In any given case, there will be a distribution
in the value of the angle existing between the directions of the flows
and the field. We have only considered the extremities of this distribution
(flows aligned or perpendicular to the mean field direction), in general
for a given angle both velocity components ($\mathbf{v}_{\Vert}$
and $\mathbf{v}_{\bot}$) would have to be simultaneously included
in the analysis. The total resulting effect observed will, in general,
reside somewhere in between what is obtained for these special cases.

It is also important to note that this phenomenon will only be observable
in regions where the motions of the neutral particles do not have
a zero mean velocity, i.e. $\left\langle \mathbf{v}^{n}\right\rangle \neq0$.
For, if they did, equations (\ref{eq:asigz}) and (\ref{eq:asigp})
would have exactly the same form and the velocity dispersions would
be the same for ions and neutrals. In other words, \emph{a} \emph{thermal
(or microturbulent) line profile would not show any manifestation
of the presence of the magnetic field}.

\subsection{Refined solution.}

It can be seen from equations (\ref{eq:avmp}), (\ref{eq:asigp})
and (\ref{eq:awrelax}) that the amplitude of the effective velocity
not only depends on the mean magnetic field strength but also on the
amount of momentum transfered during a collision. It is therefore
important to solve the problem for the more realistic cases where
the collisions are not necessarily {}``head on''. We, however, still
retain the assumption that all are perfectly elastic.

We now we get for the mean velocities and dispersions: \begin{eqnarray}
\left\langle \mathbf{v}_{\Vert}\right\rangle  & = & \left\langle \mathbf{v}_{\Vert}^{n}\right\rangle \label{eq:vz}\\
\left\langle \mathbf{v}_{\bot}\right\rangle  & = & \frac{\left\langle \mathbf{v}_{\bot}^{n}\right\rangle +\left\langle \omega_{r}\right\rangle ^{-1}\left[\left\langle \mathbf{v}_{\bot}^{n}\right\rangle \times\left\langle \overrightarrow{\omega_{g}}\right\rangle \right]}{1+\left(\frac{\left\langle \overrightarrow{\omega_{g}}\right\rangle }{\left\langle \omega_{r}\right\rangle }\right)^{2}}\label{eq:vperp}\\
\sigma_{\Vert}^{2} & = & \frac{\left[\left\langle \left|\mathbf{v}_{\bot}^{n}\right|^{2}\right\rangle -\left\langle \mathbf{v}_{\bot}\right\rangle ^{2}\right]M\left(a,\gamma\right)+\left[\sigma_{\Vert}^{n}\right]^{2}N\left(a,\gamma\right)}{\left[\frac{2\left\langle a\cos\left(\gamma\right)\right\rangle }{\left\langle a^{2}\right\rangle }-1\right]D\left(a,\gamma\right)}\label{eq:sigz}\\
\sigma_{\bot}^{2} & = & \frac{\left[\left\langle \left|\mathbf{v}_{\bot}^{n}\right|^{2}\right\rangle -\left\langle \mathbf{v}_{\bot}\right\rangle ^{2}\right]P\left(a,\gamma\right)+\left[\sigma_{\Vert}^{n}\right]^{2}Q\left(a,\gamma\right)}{\left[\frac{2\left\langle a\cos\left(\gamma\right)\right\rangle }{\left\langle a^{2}\right\rangle }-1\right]D\left(a,\gamma\right)}\label{eq:sigp}\\
\sigma_{T}^{2} & = & \frac{\left[\left\langle \left|\mathbf{v}_{\bot}^{n}\right|^{2}\right\rangle -\left\langle \mathbf{v}_{\bot}\right\rangle ^{2}\right]+\left[\sigma_{\Vert}^{n}\right]^{2}}{\left[\frac{2\left\langle a\cos\left(\gamma\right)\right\rangle }{\left\langle a^{2}\right\rangle }-1\right]}\label{eq:sigt}\end{eqnarray}

where:

\begin{eqnarray*}
D\left(a,\gamma\right) & = & 2\left\langle a\cos\left(\gamma\right)\right\rangle -\left\langle a^{2}\cos^{2}\left(\gamma\right)\right\rangle +\frac{1}{2}\left\langle a^{2}\sin^{2}\left(\gamma\right)\right\rangle \\
M\left(a,\gamma\right) & = & \frac{\left\langle a\cos\left(\gamma\right)\right\rangle }{\left\langle a^{2}\right\rangle }\left\langle a^{2}\sin^{2}\left(\gamma\right)\right\rangle \\
N\left(a,\gamma\right) & = & \left[\frac{2\left\langle a\cos\left(\gamma\right)\right\rangle }{\left\langle a^{2}\right\rangle }-1\right]\left\langle a^{2}\cos^{2}\left(\gamma\right)\right\rangle +\frac{1}{2}\left\langle a^{2}\sin^{2}\left(\gamma\right)\right\rangle \\
P\left(a,\gamma\right) & = & D\left(a,\gamma\right)-M\left(a,\gamma\right)\\
Q\left(a,\gamma\right) & = & D\left(a,\gamma\right)-N\left(a,\gamma\right)\end{eqnarray*}

The relaxation rate is now given by:

\begin{equation}
\left\langle \omega_{r}\right\rangle =\left\langle a\cos\left(\gamma\right)\right\rangle \nu_{c}\,\,.\label{eq:wrelaxation}\end{equation}

$a$ is the ratio of the change in velocity of the ion after a collision
to the initial relative velocity between the two particles, $\gamma$
is the scattering angle of the ion as measured in its initial rest
frame and the different averages ($\left\langle a\cos\left(\gamma\right)\right\rangle $,
\ldots{}) are evaluated over the space of the scattering angle of
the colliding neutral particle in the same frame of reference. We
give in table \ref{ta:avg} set of values for the different averages
for two different ion molecular masses ($A_{i}=29$ and $45$). We
would also like to point out that from numerical calculations we obtain
the following relations (which could also be derived from simple physical
considerations):

\begin{eqnarray*}
\left\langle a\cos\left(\gamma\right)\right\rangle  & \simeq & \frac{\mu}{m_{i}}\\
\frac{2\left\langle a\cos\left(\gamma\right)\right\rangle }{\left\langle a^{2}\right\rangle } & = & \frac{m_{i}}{\mu}\end{eqnarray*}

which allow us to see that our first evaluation of the relaxation
rate in equation (\ref{eq:awrelax}) was wrong by a factor of two
whereas the denominator in equation (\ref{eq:sigt}) for $\sigma_{T}^{2}$
is unchanged. Another difference between equations (\ref{eq:sigz})-(\ref{eq:sigp})
and (\ref{eq:asigz})-(\ref{eq:asigp}) is that we now have a {}``mixing''
between the two variances; the neutral dispersion parallel (perpendicular)
to the magnetic field affects the ion velocity dispersion perpendicular
(parallel) to the field. This is of course due to our more realistic
treatment of the collisions where momentum can now be transfered between
different directions. But despite these few changes, all the conclusions
reached in the previous section still hold.

\placetable{ta:avg}

\begin{figure}
\begin{center}\plotone{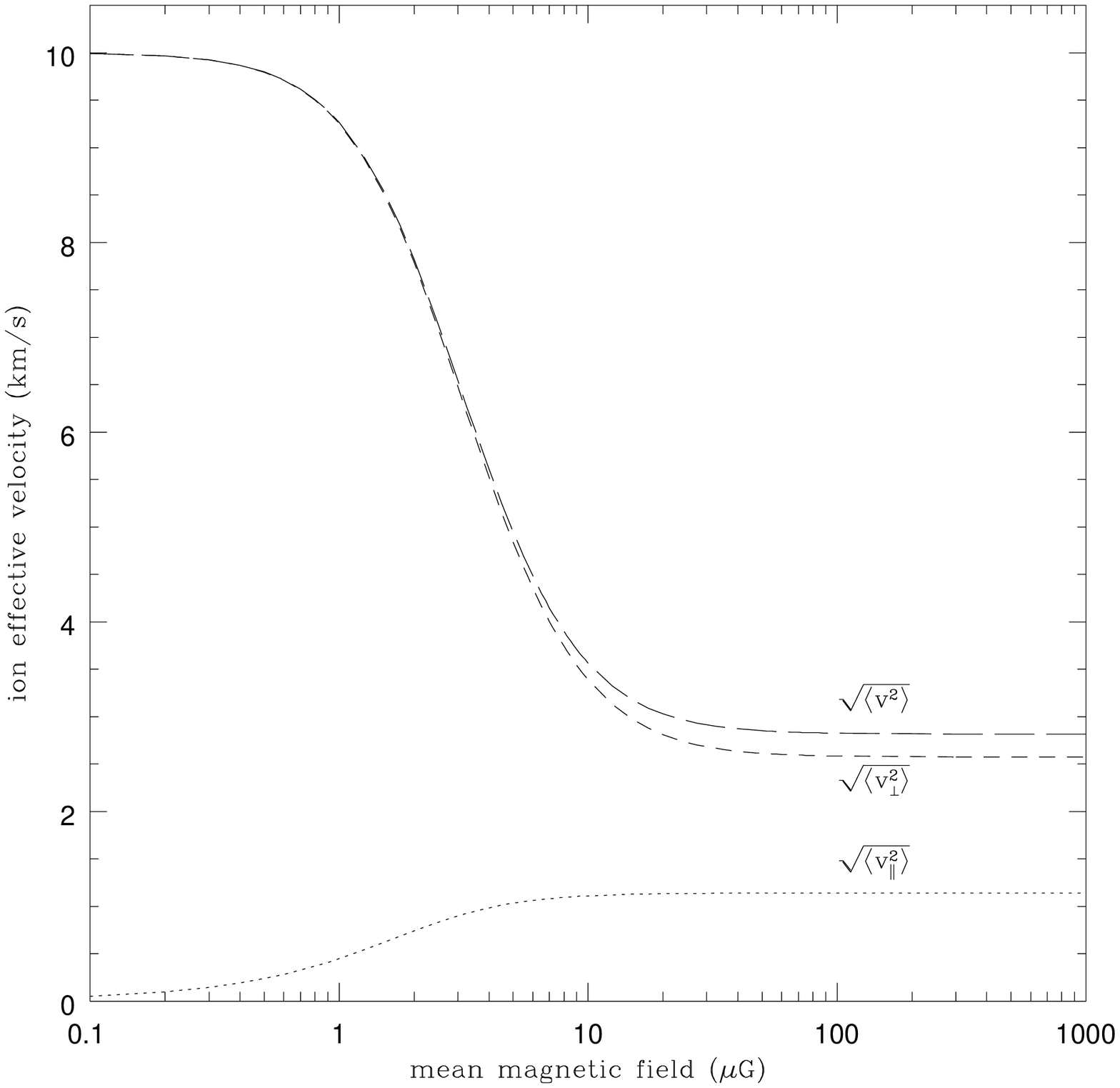}\end{center}

\caption{\label{fig:vvsB}Ion effective velocity ($\sqrt{\left\langle \mathbf{v}_{\Vert}^{2}\right\rangle }$,
$\sqrt{\left\langle \mathbf{v}_{\bot}^{2}\right\rangle }$ and $\sqrt{\left\langle \mathbf{v}^{2}\right\rangle }$)
as a function of the mean magnetic field strength when $\mathbf{v}_{\Vert}^{n}=0$,
$\left|\mathbf{v}_{\bot}^{n}\right|=10$ km/s, $n=5\times10^{6}$
cm$^{-3}$ and $A_{i}=29$.}
\end{figure}

In figure \ref{fig:vvsB}, we have plotted curves for the ion effective
velocity (according to equations (\ref{eq:vz})-(\ref{eq:wrelaxation}))
as a function of the mean magnetic field intensity when $\left\langle \mathbf{v}_{\Vert}^{n}\right\rangle =0$,
$\left|\left\langle \mathbf{v}_{\bot}^{n}\right\rangle \right|=10$
km/s and the neutral density $n=5\times10^{6}$ cm$^{-3}$. For simplicity,
we also set $\left\langle \left|\mathbf{v}^{n}\right|^{2}\right\rangle =\left\langle \mathbf{v}^{n}\right\rangle ^{2}$.
As can be seen, the field makes its presence felt even for relatively
weak intensities ($\left\langle B\right\rangle \ga10\,\mu$G) and
the transition between the regimes where ions follow the neutral flow
to where they are trapped by the field is quite abrupt ($1\,\mu$G
$\la\left\langle B\right\rangle \la10\,\mu$G). It is important to
point out that it is, in principle, possible to evaluate the mean
intensity of the magnetic field from this figure. One would have to
compare the width of neutral and ion spectra and find the field strength
that matches it at a given angle between the direction of the field
and the line of sight to the observer. When this line of sight is
parallel (perpendicular) to the field, the ion velocity should follow
the curve for $\sqrt{\left\langle \mathbf{v}_{\Vert}^{2}\right\rangle }$
($\sqrt{\left\langle \mathbf{v}_{\bot}^{2}\right\rangle }$). There
are, however, a few things that stop us from achieving this. Among
these are the unknown amount of velocity dispersion inherent to the
neutral flow and the uncertainty in the neutral density (used to determine
the collision rate). We should also point out that recent observations
indicate that field strengths in molecular clouds are in the range
of a few hundreds $\mu$G \citep{Crutcher et al. 1999} where our
curves in figure \ref{fig:vvsB} are basically flat and the ion velocity
is insensitive to field strength variations. But we should not rule
out the possibility that the determination of the field properties
could become feasible in the future when using more complete models.
Still at this point, it is nonetheless possible to assign a lower
value for the field strength (given the density of the gas).

\placefigure{fig:vvsB}

\section{Observational evidence.}

We would now like to bring support to our assertion of the previous
section that we should expect that in certain conditions ion lines
should have narrower profiles than neutral species. Furthermore, we
contend that this situation is likely to happen frequently. The reason
for this lies with equations (\ref{eq:vz})-(\ref{eq:vperp}). 

As we recall, the requirement is for a poor alignment between the
local magnetic field and the {}``mean'' flow; the effect being maximized
when they are perpendicular to each other. Assuming that the magnetic
field is strong enough, we can see from the aforementioned equations
that ions in general will not follow the motion of the flow but will
tend to drift away in a direction perpendicular to it. But since the
total mean ionic velocity is minimal when the flow and the field are
perpendicular to each other, charged particles will be more likely
to aggregate in regions where this is the case. It is therefore very
tempting to define two groups of objects (as far as the observation
of the effects of the magnetic field is concerned): 

\begin{enumerate}
\item the field is aligned with the flow, no significant differences between
the spectra of comparable neutral and ion
\item no general alignment between the field and the flow(s), the ion species
exhibit narrower line profiles and suppression of the high velocity
wings in their spectra.
\end{enumerate}
In what follows, we present observational evidence showing some examples
that we believe belong to the second class.

\subsection{Observations of HCN, HCO$^{+}$ and N$_{2}$H$^{+}$.}

In HCN, HCO$^{+}$ and N$_{2}$H$^{+}$ we have three fairly similar
molecules except for the obvious fact that the last two are ions whereas
the first is a neutral. Indeed, they are all linear molecules with
comparable atomic mass, similar rotation spectra and almost identical
critical densities ($n_{c}\simeq10^{6}$ cm$^{-3}$). They therefore
appear to be excellent candidates to test our proposal.

\begin{figure}
\notetoeditor{the four EPS files should appear in the figure fig:spectra as shown with the commands below}

\begin{center}\epsscale{0.9}\plottwo{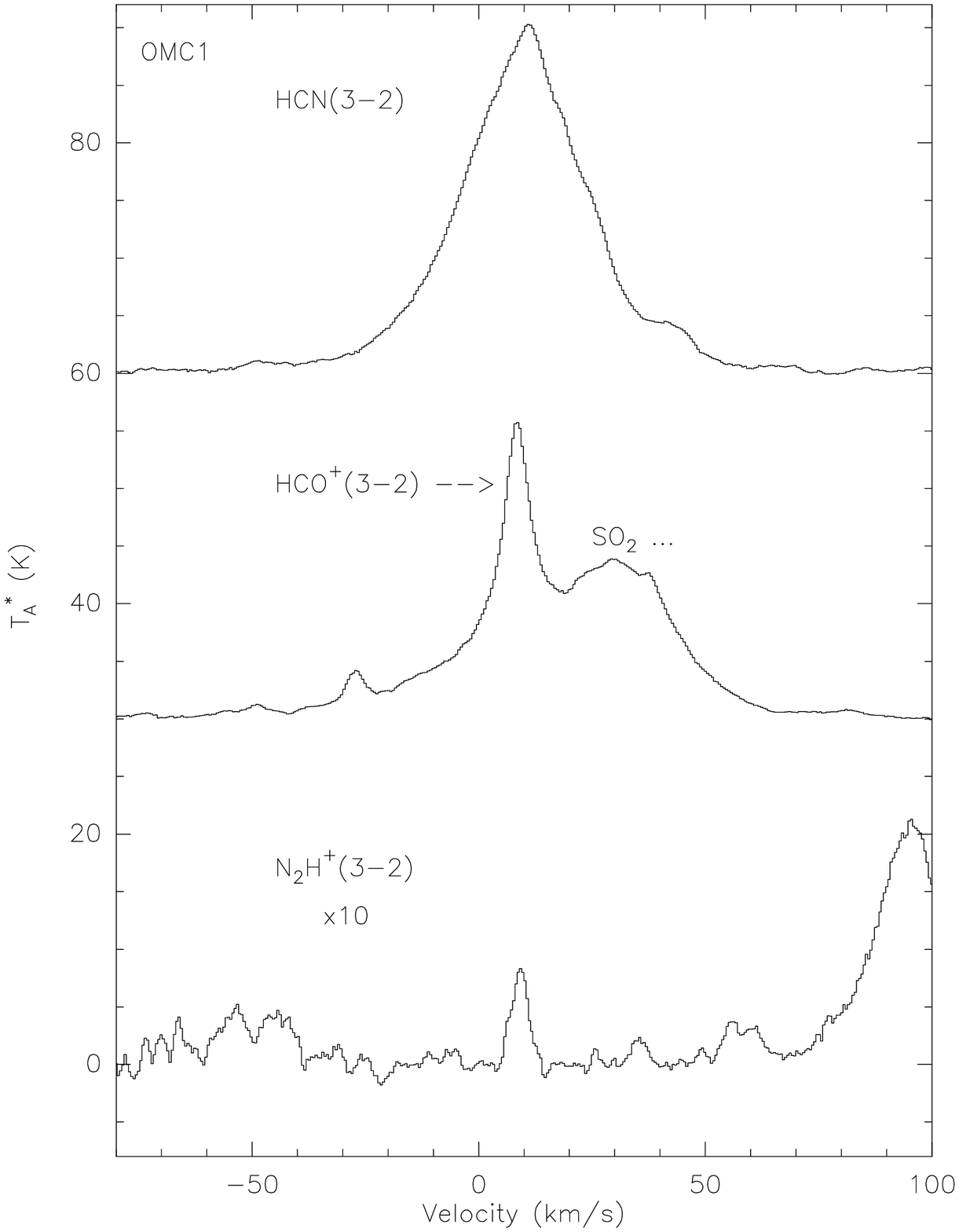}{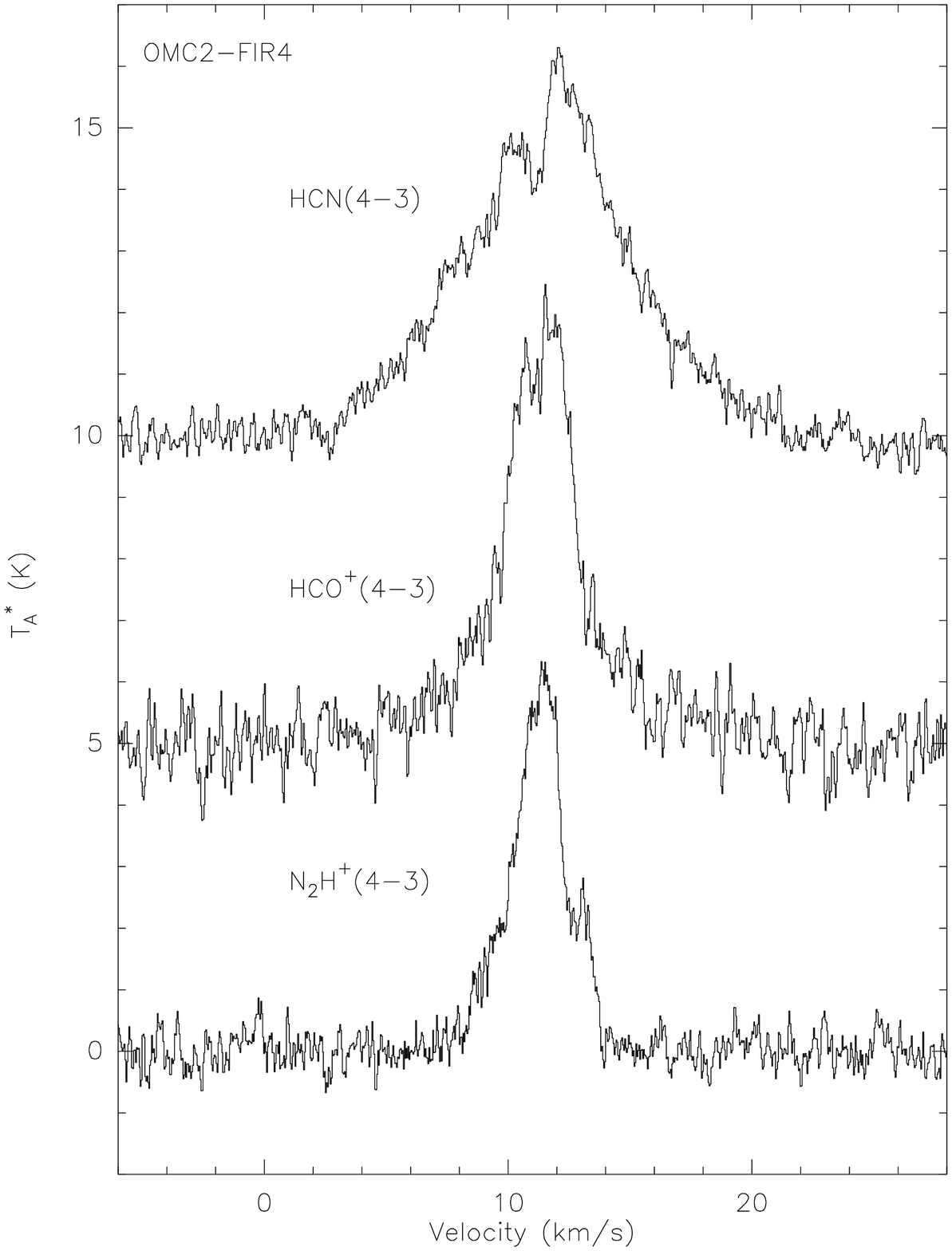}\end{center}

\begin{center}\epsscale{0.9}\plottwo{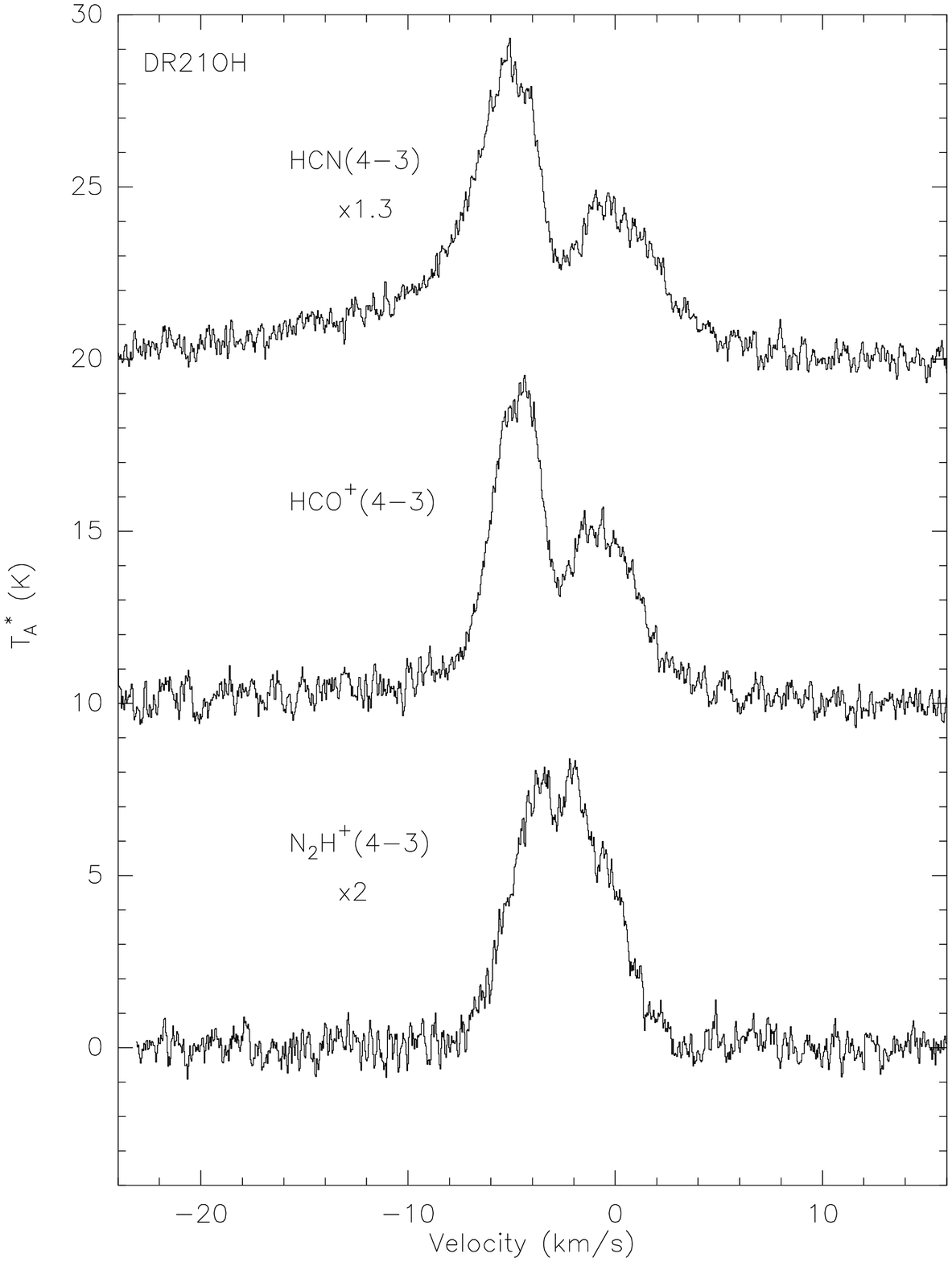}{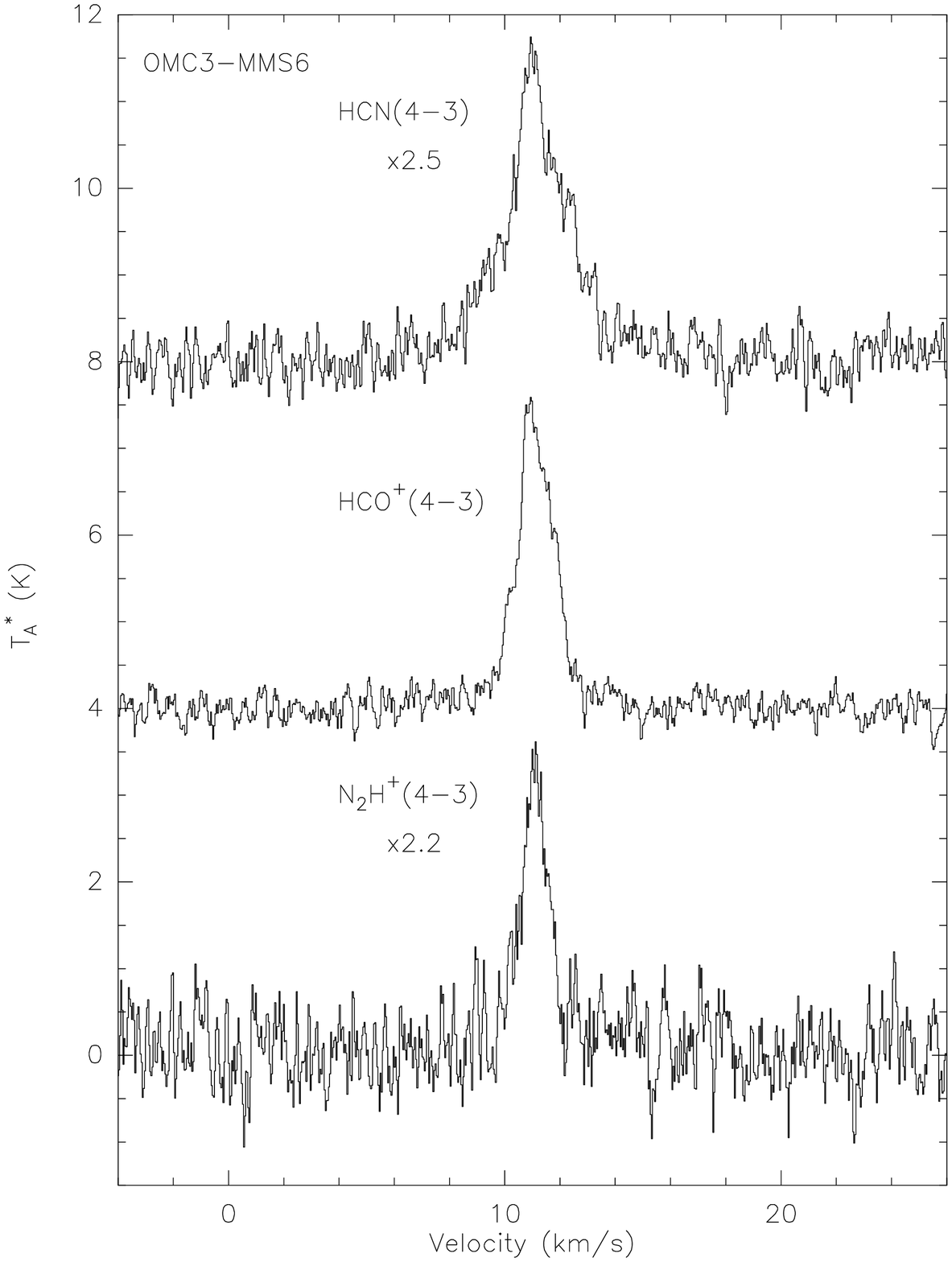}\end{center}

\caption{\label{fig:spectra}HCN (top), HCO$^{+}$(middle) and N$_{2}$H$^{+}$
(bottom) observations at the position of peak intensity of (clockwise
starting from top left): OMC1, OMC2-FIR4, OMC3-MMS6 and DR21OH. The
bump at $\sim30$ km/s in the middle spectrum of OMC1 is a contamination
from other species (SO$_{2}$, $^{13}$CH$_{3}$CN, \ldots{}).}
\end{figure}

\begin{figure}
\notetoeditor{the four EPS files should appear in the figure fig:OMC2 as shown with the commands below}

\resizebox*{16cm}{10cm}{\rotatebox{270}{\includegraphics{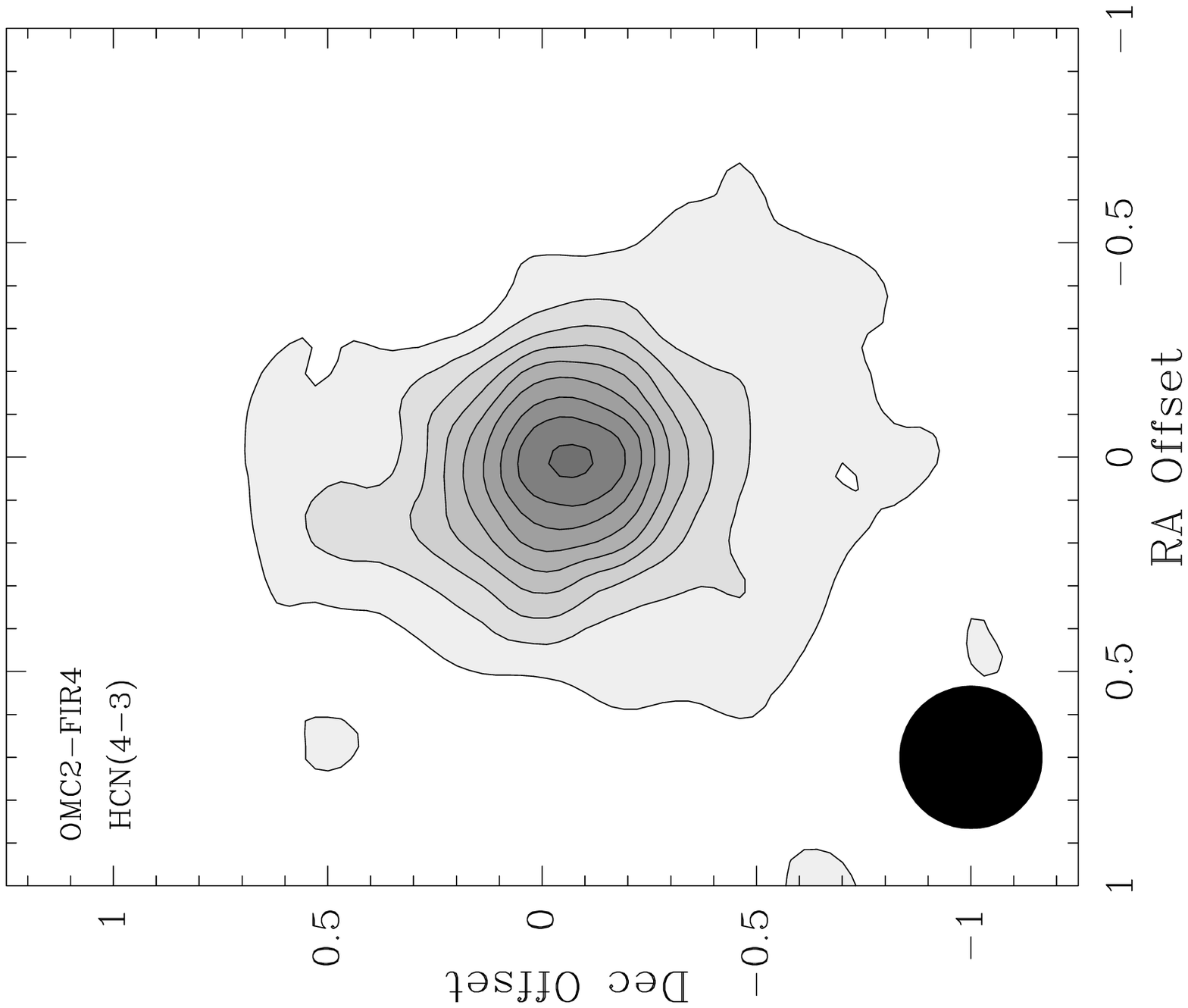}}\hspace{0.5cm}\rotatebox{270}{\includegraphics{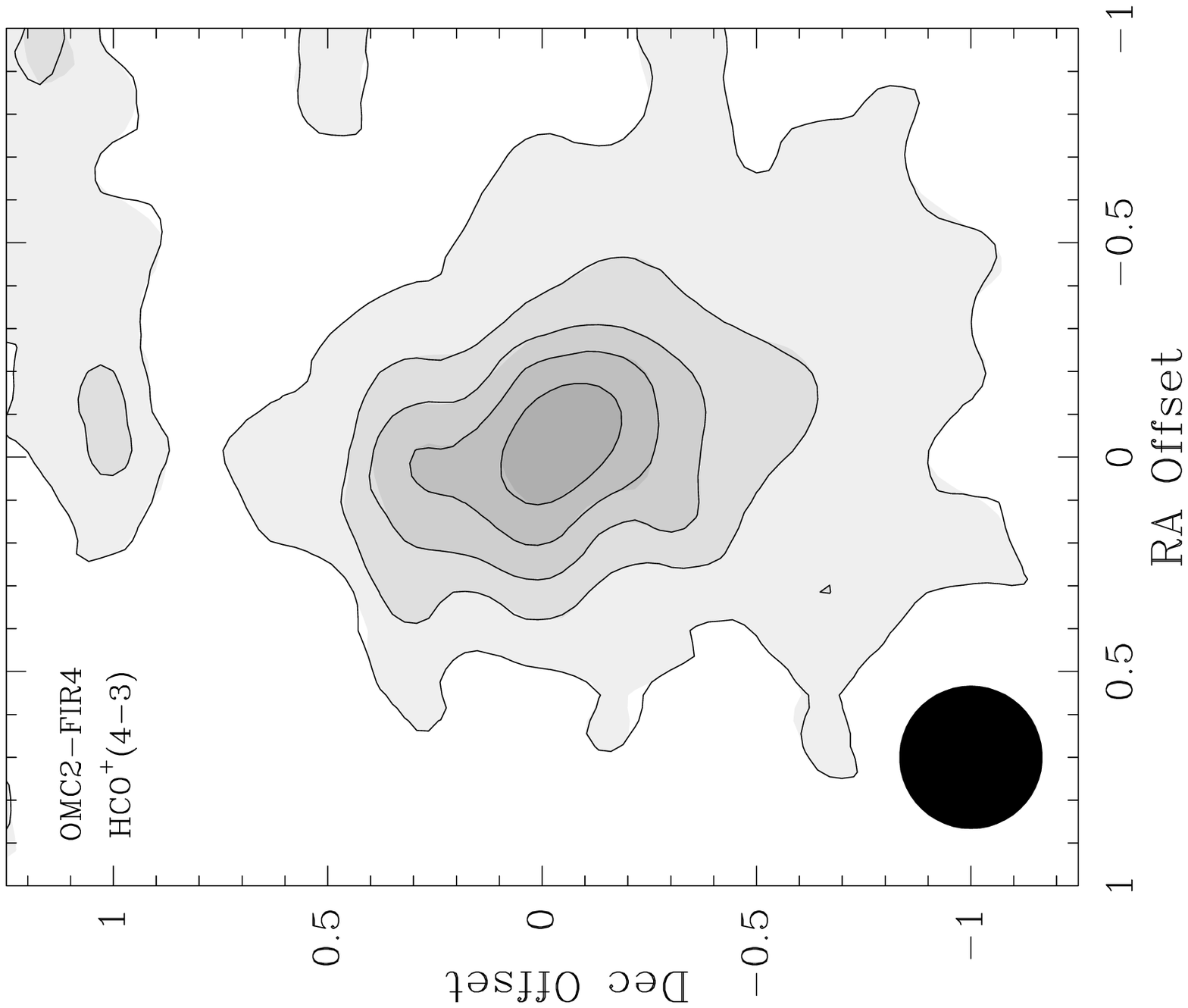}}}

\vspace{1cm}

\caption{\label{fig:OMC2}HCN and HCO$^{+}$ ($J\rightarrow4-3$) maps of
OMC2-FIR4. Although the ion spatial distribution appears somewhat
more extended, the two peaks are well aligned and the HCN and HCO$^{+}$
systematic velocities agree (from figure \ref{fig:spectra}). The
two maps have the same lowest contour level of 5 K$\cdot$km/s ($\sim10\sigma$)
and the following levels increase linearly with an interval of 5 K$\cdot$km/s.
The maps' grid spacing of $10\arcsec$ is half of the beam width (shown
in the lower left corners). The pointing accuracy is better than $\sim5\arcsec$.}
\end{figure}

We present in figure \ref{fig:spectra} spectra of these three molecules
in the $J\rightarrow3-2$ transition for OMC1 and $J\rightarrow4-3$
for OMC2-FIR4, OMC3-MMS6 and DR21OH obtained at the Caltech Submillimeter
Observatory with the 200-300 GHz and 300-400 GHz receivers. The observations
(including the maps presented in figure \ref{fig:OMC2}) were done
on several nights during the months of March, April, June and August
1999. Pointing was checked at regular interval using scans made on
planets available at the time. Telescope efficiencies were calculated
to be $\sim70$ \% for the 200-300 GHz receiver (beam width of $\sim32\arcsec$)
and $\sim60$ \% for the 300-400 GHz receiver (beam width of $\sim20\arcsec$).

As can be seen from figure \ref{fig:spectra}, all these spectra show
the characteristics described earlier; the ion lines are all significantly
narrower than the HCN lines and the suppression of the wings is also
obvious. These conclusions are made more definite in table \ref{ta:widths}
where we present a comparison of the line widths (more precisely the
standard deviations $\sigma_{v}$) of the species for these molecular
clouds. The widths were measured after the lines were modeled with
a multi-Gaussian profile. We interpret the fact that the predictions
made here on the differences in the appearance in the spectra between
comparable neutral and ion molecular species seem to be easily observable
in molecular clouds as strong evidence in favor of our assertions.

We should however point out that the differences between the line
profiles of neutral and ion species could be partly due to other factors.
Indeed, the assumed coexistence of the different molecular species
is not necessarily correct. Although we tried to choose ion and neutral
species that are as similar as possible, it is likely that their observations
sample different parts of the molecular clouds. In some cases, it
might even be possible that two species are exposed to significantly
different dynamical processes that could cause one of them to exhibit
a narrower or larger line profile no matter what the effect of the
magnetic field might be. Maps of the different molecular species for
a given object could shed some light on this question. In the case
of OMC1, \citet{Ungerechts 1997} have extensively mapped the Orion
molecular cloud (around OMC1) with 20 different chemical and isotopic
molecular species. Amongst other things, they find an impressive degree
of uniformity in the chemical abundances; but most interesting to
us is how similar their maps of HCN and HCO$^{+}$ are ($J\rightarrow1-0$,
beam width of $\sim50\arcsec$). On the other hand, N$_{2}$H$^{+}$
seems to have a very different spatial distribution (see their figure
2). We also present in figure \ref{fig:OMC2} our maps of OMC2-FIR4
in HCN and HCO$^{+}$ ($J\rightarrow4-3$) done at the CSO. Although
the ion species appears somewhat more extended, the two peaks are
well aligned. This and the fact the two species have similar line
center velocity (see figure \ref{fig:spectra}) brings support to
the coexistence assumption for these molecules in the OMC2 cloud.

The question of abundances is also important. For example, a given
species could exhibit more opaque lines which could change some characteristics
of the line profiles (such as the relative importance of the high
velocity wings, saturation or self-absorption). In fact, we believe
that the differences between the spectra of N$_{2}$H$^{+}$ and HCO$^{+}$
are probably the result of this effect (see the case of DR21OH in
figure \ref{fig:spectra}, the self-absorption feature at $\sim-2.5$
km/s is much weaker in the N$_{2}$H$^{+}$ spectrum implying a lower
abundance).

Chemical differentiation could also play a role in explaining the
differences between spectra. In this effect, we expect that our analysis
only applies to long-lived ion species (like the protonated molecules
considered here), other more reactive species (CO$^{+}$, SO$^{+}$,
\ldots{}) should more or less behave as neutrals as almost every
collision in which they are involved would entail a chemical reaction
\citep{Schilke 1999}.

\placetable{ta:widths}

\placefigure{fig:spectra}

\placefigure{fig:OMC2}

\section{Relaxation time and interaction between charged particles.}

In our analysis of the problem in section \ref{sec:analysis}, we
have considered the behavior of a given ion without taking into account
the presence of any other charged particles. One might intuitively
guess that for a weakly ionized plasma, like the ones probed with
our observations at the CSO presented in the last section, it is probably
safe to do so.

One can make sure of this by comparing the mean time between collisions
for charged particles in such a plasma with the time it takes an ion
to relax to its steady state after being excited to a different velocity.
The excitation could occur, for example, during a collision with an
electron. However, because of the huge disparity between the masses
of the two particles, it is unlikely that the ion would be much perturbed
by such an encounter. A collision with another ion is more likely
to produce a significant change in its velocity.

At any rate, it turns out that the ion relaxation time is given by
the reciprocal of the relaxation rate defined by equation (\ref{eq:wrelaxation}),
as one can verify when solving the equation of motion (\ref{eq:motion})
for a mean collision force $\left\langle \mathbf{F}_{c}\right\rangle $.
The solution shows that the ion velocity decays exponentially with
a time constant equal to $\left\langle \omega_{r}\right\rangle ^{-1}$.
When this time is compared to the mean collision time between ions,
it is seen to be at least a few orders of magnitude smaller. Our simplified
analysis is therefore justified.

\section{Conclusion.}

We argue that the presence of a magnetic field in a weakly ionized
plasma can be easily detected through a comparison of ion and neutral
line profiles. More precisely, we expect ion lines to often exhibit
narrower profiles and significant suppression of high velocity wings.
We have presented observational evidence obtained for four different
molecular clouds which agrees with our theory. 

Because of the low intensities of field required for the effect to
be noticeable, we expect the phenomenon to be widespread. However,
where the mean magnetic field is aligned with the neutral flow(s),
or where the line width is dominated by the thermal width, should
not show any significant differences between comparable neutral and
ion spectra. These aspects of the problem will be treated in subsequent
papers.

\acknowledgements{We thank Prof. J. Zmuidzinas for his help during the preparation
of this paper. M. Houde's work was done in part with the assistance
of grants from FCAR and the Département de Physique of the Université
de Montréal. The Caltech Submillimeter Observatory is funded by the
NSF through contract AST 9615025. }

\appendix

\section{Derivations.}

Equations (\ref{eq:avmz}) and (\ref{eq:avmp}) can be easily derived
by taking the mean of the equation of motion (\ref{eq:motion}) and
then assuming a steady state:

\[
\left\langle \frac{d\mathbf{v}}{dt}\right\rangle =0\]

Using equation (\ref{eq:force}) as an approximation for the force
of interaction between ions and neutrals we get from the mean equation
of motion:

\begin{eqnarray}
\left\langle \mathbf{F}_{c\Vert}\right\rangle  & = & 0\label{eq:mFcp}\\
\left\langle \mathbf{F}_{c\bot}\right\rangle  & = & -\frac{e}{m_{i}c}\left\langle \mathbf{v}_{\bot}\right\rangle \times\left\langle \mathbf{B}\right\rangle \label{eq:mFcn}\end{eqnarray}

with 

\begin{equation}
\left\langle \mathbf{F}_{c}\right\rangle =-2\frac{\mu}{m_{i}}\nu_{c}\left(\left\langle \mathbf{v}\right\rangle -\left\langle \mathbf{v}^{n}\right\rangle \right)\label{eq:mFc}\end{equation}

where $\nu_{c}$ is the mean collision rate. From equation (\ref{eq:mFcp})
and (\ref{eq:mFc}) it is straightforward to get equation (\ref{eq:avmz})
for $\left\langle \mathbf{v}_{\Vert}\right\rangle $. Equation (\ref{eq:mFcn})
can be transformed to:

\[
\left\langle \mathbf{v}_{\bot}\right\rangle =-\frac{\left\langle \omega_{r}\right\rangle }{\left\langle \overrightarrow{\omega_{g}}\right\rangle ^{2}}\left[\left(\left\langle \mathbf{v}_{\bot}\right\rangle -\left\langle \mathbf{v}_{\bot}^{n}\right\rangle \right)\times\left\langle \overrightarrow{\omega_{g}}\right\rangle \right]\]

into which we can insert equation (\ref{eq:mFcn}) using (\ref{eq:mFc})
and finally get equation (\ref{eq:avmp}) which we rewrite here:

\begin{equation}
\left\langle \mathbf{v}_{\bot}\right\rangle =\frac{\left\langle \mathbf{v}_{\bot}^{n}\right\rangle +\left\langle \omega_{r}\right\rangle ^{-1}\left[\left\langle \mathbf{v}_{\bot}^{n}\right\rangle \times\left\langle \overrightarrow{\omega_{g}}\right\rangle \right]}{1+\left(\frac{\left\langle \overrightarrow{\omega_{g}}\right\rangle }{\left\langle \omega_{r}\right\rangle }\right)^{2}}\label{eq:apavmp}\end{equation}

$\left\langle \overrightarrow{\omega_{g}}\right\rangle $ and $\left\langle \omega_{r}\right\rangle $
are defined in equation (\ref{eq:gyrofreq}) and (\ref{eq:awrelax})
respectively.

Equations (\ref{eq:asigz}) and (\ref{eq:asigp}) for the velocity
dispersions can be derived by using the following procedure. We concentrate
on the equation for $\sigma_{\bot}^{2}$. If we assume that the time
of interaction during a collision is infinitively small, we can express
$\mathbf{v}_{\bot}'$, the ion velocity (perpendicular to the field)
immediately after a collision, as a function of the velocity $\mathbf{v}_{\bot}$
just before the collision:

\begin{equation}
\mathbf{v}_{\bot}'=\mathbf{v}_{\bot}+\frac{\Delta\mathbf{p}_{\bot}}{m_{i}}\label{eq:vprime}\end{equation}

with $\Delta\mathbf{p}_{\bot}=-2\mu\left(\mathbf{v}_{\bot}-\mathbf{v}_{\bot}^{n}\right)$
the change in the ion momentum. Upon taking the mean of the square
of equation (\ref{eq:vprime}) while once again imposing steady state
conditions, i.e. $\left\langle \left|\mathbf{v}_{\bot}'\right|^{2}\right\rangle =\left\langle \left|\mathbf{v}_{\bot}\right|^{2}\right\rangle $,
we are left with:

\begin{equation}
\left\langle \mathbf{v}_{\bot}\cdot\frac{\Delta\mathbf{p}_{\bot}}{m_{i}}\right\rangle =-\frac{1}{2}\left\langle \left|\frac{\Delta\mathbf{p}_{\bot}}{m_{i}}\right|^{2}\right\rangle \label{eq:scpro}\end{equation}

Taking into account the fact that $\left\langle \mathbf{v}_{\bot}\right\rangle \cdot\left\langle \mathbf{v}_{\bot}^{n}\right\rangle =\left\langle \mathbf{v}_{\bot}\right\rangle ^{2}$,
as can be verified with equation (\ref{eq:apavmp}), it is then straightforward
to obtain:

\begin{equation}
\sigma_{\bot}^{2}=\frac{\left\langle \left|\mathbf{v}_{\bot}^{n}\right|^{2}\right\rangle -\left\langle \mathbf{v}_{\bot}\right\rangle ^{2}}{\left[\frac{m_{i}}{\mu}-1\right]}\label{eq:apasigp}\end{equation}

The equation for $\sigma_{\Vert}^{2}$ follows from the same procedure.

The more refined set of equations (\ref{eq:vz})-(\ref{eq:wrelaxation})
can also be obtained in similar fashion but the task is rendered somewhat
more complicated by the fact the interaction force $\mathbf{F}_{c}$
is not simply given by equation (\ref{eq:force}). We will omit their
derivation here as they are somewhat lengthy and don't bring anything
substantially new to the discussion.

One last word concerning the energetics involved in the collision
process. From the above discussion, it is apparent that an ion which
starts off with the same velocity as the neutral flow would eventually
settle into a steady state of lower kinetic energy (assuming a strong
magnetic field). Once again focusing on the approximate head-on collision
model as seen in the reference frame of the observer, we can understand
this loss of energy by noting that the drag force felt by the ion
(equation (\ref{eq:force})) is proportional to the relative velocity
between the colliding particles. As the ion is attempting to travel
in a circular orbit around a given guiding center, this relative velocity
is greater when it is going upstream (and losing energy through collisions)
than when it is going downstream (and gaining energy from the collisions).
This will lead, on average, to a net loss of energy over a complete
orbit.

\pagebreak

\begin{table}
\begin{center}\begin{tabular}{lll}
\hline 
Parameter&
$A_{i}=29$&
$A_{1}=45$\tabularnewline
\hline 
$\left\langle a\cos\left(\gamma\right)\right\rangle $&
0.07734&
0.05028\tabularnewline
$\left\langle a^{2}\cos^{2}\left(\gamma\right)\right\rangle $&
0.00778&
0.00332\tabularnewline
$\left\langle a^{2}\sin^{2}\left(\gamma\right)\right\rangle $&
0.00359&
0.00157\tabularnewline
$\left\langle a^{2}\right\rangle $&
0.01137&
0.00489\tabularnewline
\hline
\end{tabular}\end{center}

\caption{\label{ta:avg}Numerical values for the different averages present
in equations (\ref{eq:vz})-(\ref{eq:wrelaxation}) for ion molecular
masses of 29 and 45.}
\end{table}

\begin{table}
\begin{center}\begin{tabular}{l|cc|ccc}
\hline 
&
\multicolumn{2}{c|}{Coordinates (1950)}&
\multicolumn{3}{c}{$\sigma_{v}$(km/s)}\tabularnewline
\cline{2-3} 
\hline 
Object&
R.A.&
Decl.&
HCN&
HCO$^{+}$&
N$_{2}$H$^{+}$\tabularnewline
\hline 
OMC-1&
5 32 47.2&
-05 24 25.3&
17.42&
3.23&
1.87\tabularnewline
OMC-2 FIR 4&
5 32 59.0&
-05 11 54.0&
3.59&
2.72&
1.34\tabularnewline
OMC-3 MMS 6&
5 32 55.6&
-05 03 25.0&
1.40&
0.71&
0.56\tabularnewline
DR 21(OH)&
20 37 13.0&
42 12 00.0&
5.75&
4.61&
2.08\tabularnewline
\hline
\end{tabular}\end{center}

\caption{\label{ta:widths}Comparisons of line widths (standard deviations
$\sigma_{v}$) between HCN, HCO$^{+}$, and N$_{2}$H$^{+}$ for the
four objects presented in figure \ref{fig:spectra}.}
\end{table}

\end{document}